\documentclass[conference]{IEEEtran}
\IEEEoverridecommandlockouts
\usepackage{cite}
\usepackage{amsmath,amssymb,amsfonts}
\usepackage{algorithmic}
\usepackage{graphicx}
\usepackage{textcomp}
\usepackage{xcolor}
\usepackage{booktabs}
\usepackage{siunitx}
\usepackage{subcaption}
 \usepackage{booktabs}

\usepackage{amsthm}
\usepackage{hyperref}
\usepackage{float}

\theoremstyle{definition}

\begingroup\expandafter\expandafter\expandafter\endgroup
\expandafter\ifx\csname pdfsuppresswarningpagegroup\endcsname\relax
\else
\fi

\def\BibTeX{{\rm B\kern-.05em{\sc i\kern-.025em b}\kern-.08em
    T\kern-.1667em\lower.7ex\hbox{E}\kern-.125emX}}

\begin{document}

\title{Quality of service based radar resource management for synchronisation problems
\thanks{The ﬁrst two authors contributed equally to this work.}
}

\author{
	\IEEEauthorblockN{Tobias M\"uller, Sebastian Durst, Pascal Marquardt and Stefan Br\"uggenwirth}
	\IEEEauthorblockA{\textit{Fraunhofer-Institut f\"ur Hochfrequenzphysik und Radartechnik FHR}\\
		Wachtberg, Germany \\
		tobias.mueller@fhr.fraunhofer.de, sebastian.durst@fhr.fraunhofer.de\\
		+49 228 9435-79019
	}
}

\maketitle

\begin{abstract}
	An intelligent radar resource management is an essential building block of any modern radar system.
	The quality of service based resource allocation model (Q-RAM) provides a framework for profound and quantifiable decision-making but
	lacks a representation of inter-task dependencies that can e.g.\ arise for tracking and synchronisation tasks.
	As a consequence, synchronisation is usually performed in fixed non-optimal patterns.
	We present an extension of Q-RAM which enables the resource allocation to consider complex inter-task dependencies and can produce adaptive and intelligent synchronisation schemes.
	The provided experimental results demonstrate a significant improvement over traditional strategies.
\end{abstract}



\section{Introduction}
\label{sec:intro}

Resource management (i.e.\ task prioritisation, resource allocation and scheduling) is an important part of any modern radar system, as these have to simultaneously perform potentially conflicting functions,
and even more so of any cognitive setup.
One important role of the resource management is to select operational parameters from a multitude of possible task configurations differing in resource requirements and resulting utility with the goal to optimise the overall system performance under resource constraints.
A mathematical framework describing this problem is known as \emph{Quality of service based resource allocation model (Q-RAM)}
\cite{Raj1997, Ghosh2006}.

Bi- or multi-static radar has grown in popularity because of several reasons among which are the prevention of dead times, a possibly covert operation of the receiver and a higher resilience in a military context.
In order for a multi-static radar to function, the receivers need to know the location of the transmitters and all sensors have to share a common time and frequency standard.
This is essential for accurate range and velocity estimations.
Hence, the sensitive local oscillators used in the individual sensors need to be kept synchronised \cite{Yang2011}.
Clocks being out of sync lead to higher measurement errors and thus to a degradation of the overall system performance.
These effects need to be taken into account by the resource manager to guarantee an optimal resource allocation and develop intelligent synchronisation schedules.
This is particularly true for modern and future multi-function systems, where various functions are executed on the same aperture, i.e.\ the synchronisation process itself competes with other tasks for shared resources. In many cases GNSS-based synchronisation is assumed with multi-static range measurements \cite{Prager20}.
However, this leads to unwanted dependencies especially in hostile environments and thus a separate independent GNSS-free synchronisation method is preferable in some contexts.

The standard Q-RAM framework is not able to model the effects of synchronisation appropriately and thus does not allow for a sophisticated decision-making. We propose a modification of the framework to make its use suitable in multi-static systems that require dedicated synchronisation. Furthermore, we provide experimental results that highlight a significant improvement over traditional strategies.

It is worth noting that the presented algorithm is applicable in settings other than radar synchronisation when regular calibration is required to ensure quality levels of other tasks. In particular, this is the case for IMU calibration or also channel sensing in automotive setups.

The paper is organised as follows.
The Q-RAM problem and its classical solution are briefly discussed in Sections~\ref{sec:qram_problem} and~\ref{sec:classical_solution}.
Section~\ref{sec:architecture} introduces the proposed extended framework and algorithm to enhance Q-RAM with synchronisation capabilities.
The experimental verification is presented in Section~\ref{sec:sim}. This includes a detailed description of the simulation environment with radar and target specifications as well as of the assumed synchronisation and measurement characteristics.
Finally, the paper is concluded in Section~\ref{sec:conclusion}.

\section{The Q-RAM problem}
\label{sec:qram_problem}

We will briefly introduce the Q-RAM problem in this section.
The goal of the resource allocation module in a multifunction radar is to maximise the 
\emph{utility} of a set of radar tasks by selecting \emph{operational parameters} (e.g.\ waveform, dwell period or the choice of tracking filter)
while adhering to certain \emph{resource constraints} (e.g.\ radar bandwidth, power) and taking into account the \emph{environmental conditions}, i.e.\ situational data.
A specific choice of operational parameters together with the environmental conditions determine the (expected) \emph{quality} of a task, which is usually task-type related and allows for easier, interpretable user control. Quality and situational data then define the task utility.
For $\{\tau_1,\ldots,\tau_n\}$ a set of radar tasks, $k$ types of resources with resource bounds $R_1,\ldots,R_k$ and environmental conditions $e$, the problem can be formulated as
\begin{align}
	\max_{\phi = (\phi_1,\ldots,\phi_n)}& u(\phi, e)\\
	\textrm{s.t. } \forall j=1,\ldots,k\;\;& \sum_{i=1}^n \big(g_i(\phi_i)\big)_j \leq R_j,
\end{align}
where $\phi_i$ is a configuration for task $\tau_i$, $u$ the system utility and $g_i$ functions mapping task configurations to their resource requirements
(see \cite{qram_rl} for a more detailed description).

\section{The classical approach}
\label{sec:classical_solution}

In this section we will briefly describe the approximative solution strategy to the Q-RAM problem proposed
in \cite{Raj1997, Ghosh2006}.
A notable alternative approach is given in \cite{CharlishPhD}.
First, all possible task configurations are generated and evaluated on a per task basis, i.e.\ their resulting utility and resource requirement are computed. In case of multiple resources, a scalar proxy, called \emph{compound resource}, is used.
Then a subset of configurations with a high utility for various resource levels is pre-selected and stored
in a list ordered by increasing compound resource requirement, which we will refer to as \emph{job list}.
The job lists of all tasks are given to a global optimiser, which iteratively allocates resources to the task offering the best utility-to-resource-ratio provided sufficient resources are available.
If a task is selected and assigned additional resources, we say that the task is \emph{upgraded}.
After resource allocation, the selected jobs (i.e.\ tasks with a chosen configuration) are scheduled by a scheduler.

\section{Proposed framework and algorithm}
\label{sec:architecture}

\begin{figure}[b] 
	\centering
	\includegraphics[width=.25\textwidth]{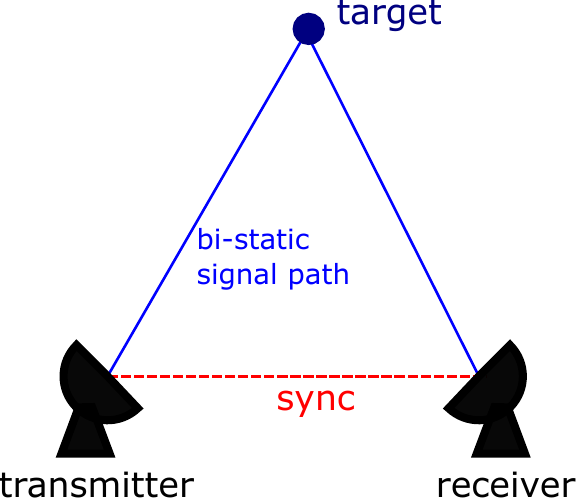}
	\caption{The bistatic setup.}
	\label{fig:setup}
\end{figure}

\begin{figure}[htb] 
	\centering
	\includegraphics[width=.45\textwidth]{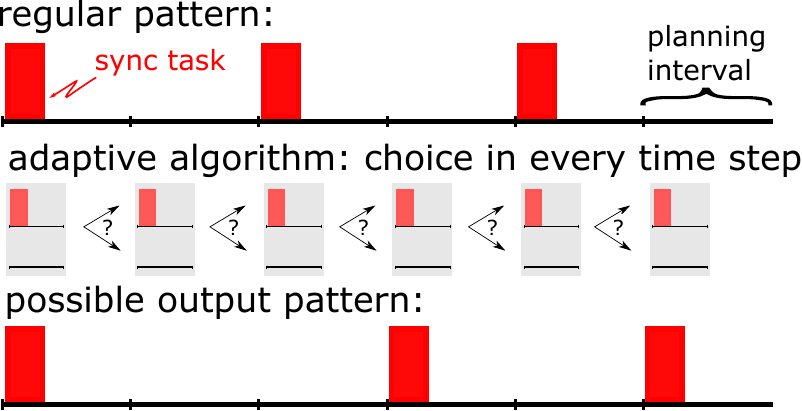}
	\caption{Regular vs.\ adaptive synchronisation scheme.} \label{fig:algo_idea}
\end{figure}

The classical Q-RAM scheme lacks the possibility to model complex inter-task dependencies, i.e.\ the quality of a task cannot depend on the expected quality of the chosen configuration for another task in the same planning period. However, these dependencies are present and important in synchronisation and calibration scenarios. More concretely, the synchronisation quality heavily influences the quality of tracking tasks as it changes the measurement accuracy. Additionally, synchronisation tasks are often considered to not have their own inherent utility but contribute to the system utility by increasing the quality of other tasks.
Clearly, there is an easy way to nevertheless perform Q-RAM in these situations. Namely, to fall back to a fixed synchronisation schedule and thus to not optimise the synchronisation itself (cf.\ top row of Figure~\ref{fig:algo_idea}).
Not surprisingly, this cannot lead to optimal resource allocations in general.

We present an extension of Q-RAM which enables the resource manager to provide adaptive and intelligent synchronisation schemes.
To this end, consider a generic bistatic radar system as depicted in Figure~\ref{fig:setup}. The transmitter has to send communication signals to the receiver on a regular basis to establish a GNSS-free time synchronisation. This is time-consuming and blocks the aperture for other tasks. Hence, it should be performed only when necessary but often enough to not degrade the quality of the other tasks.
Assume that there are multiple different synchronisation schemes $s_1,\ldots,s_n$ at hand for a fixed planning period. Each of these possesses a respective resource requirement $r_1,\ldots,r_n$ for synchronisation. The aim is to pick the scheme with the highest overall system utility. In the simplest form, this is the binary decision of performing ($s_1$) or not ($s_2$) a synchronisation task in a given planning period, in which case $r_2 = 0$. For a given $s_i$, we compute synchronisation quality and its impact on the expected utility of the other tasks which allows for a non-trivial redesign of the utility functions.
Using these and reducing the available resource budget by $r_i$, the Q-RAM algorithm to allocate resources can then be performed as usual.
This is done for all $i$ and for a given planning period, the scheme $s_i$ with the highest overall system utility is picked. In the next planning period, the process is repeated. This way a dynamic synchronisation pattern is established, that adaptively takes into account the presence and nature of other tasks to be performed as well as the current environment. The process for a binary choice is depicted in Figure~\ref{fig:algo_idea}.

Note that since the Q-RAM optimisations corresponding to the different synchronisation schemes are independent, the proposed algorithm does not increase computation time when executed in parallel.

\section{Experimental verification}
\label{sec:sim}

\begin{figure}[b]
	\centering
	\includegraphics[width=.35\textwidth]{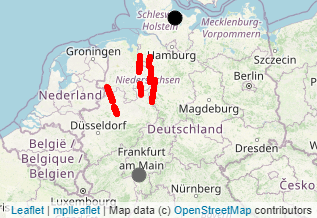}
	\caption{Scenario overview. Transmitter (grey), receiver (black) and target trajectories (red) are shown.}
	\label{fig:scenario}
\end{figure}

We demonstrate the effectiveness of the proposed algorithm via simulation in the following scenario.
The radar system is bi-static consisting of a transmitter and a receiver and requiring active time synchronisation. An increasing number of targets fly through its field of view over a period of 60 seconds. The goal of the system is to acquire and track these targets.
The positions of radar and targets are shown in Figure~\ref{fig:scenario}.
For validation, we compare our algorithm with Q-RAM for various fixed synchronisation schedules as well as with a classical rule-based scheme via a Monte Carlo simulation using Fraunhofer FHR's Cognitive Radar Simulator (CoRaSi).

\subsection{Radar and target specifications}

The active radar is assumed to be equipped with a fixed AESA antenna
reaching a maximum range of \SI{350}{\km} and focuses its emitted energy into a search fence with an opening angle of $\pm$\SI{40}{\degree} in azimuth and \SI{5}{\degree} in elevation.
The targets move with velocities of \SIrange{800}{1300}{\m \per \s} through the search fence.
The SNR for a target in the centre of the search beam with an RCS of \SI{1}{\m \squared} at a range of \SI{300}{\km} is \SI{10}{\dB}.
Without any other load, the radar needs \SI{10}{\s} for a full update of the search fence.
The direct transmission line to the passive system is attenuated (e.g.\ by terrain) with \SI{120}{\dB} such that a long integration time is necessary to synchronise both systems.
The receiver uses DBF to be independent of the transmitter.

The following (bistatic) radar equation is used to calculate the received signal-to-noise ratio
\begin{align}
	\frac{P_{\textrm{tx}} I c G_{\textrm{tx}} G_{\textrm{rx}} \sigma \lambda^2 L}{(4\pi)^3 R_{\textrm{tx}}^2 R_{\textrm{rx}}^2 k T B n_f}
\end{align}
where $P_{\textrm{tx}}$ denotes the peak power, $I$ the number of integrated pulses, $c$ the compression ratio, $G_{\textrm{tx}}$ and $G_{\textrm{tx}}$ are the gain values on transmit and receive, respectively, $\sigma$ is the radar cross-section, $\lambda$ the wavelength, $n_f$ the noise factor, $L$ system losses and $k T B$ the thermal noise.
The ranges $R_{\textrm{tx}}$ and $R_{\textrm{rx}}$ give the distance from the target at position $p \in \mathbb{R}^3$ to the transmitter and receiver, respectively. Note that the gains also depend at least indirectly on the target position $p$ via the array factor and electronic beam steering.
In case that the radar equation is evaluated for the receiver, $L$ integrates additional losses of \SI{3}{\dB} caused by the filter mismatch due to the communication part of the radar waveform.

\subsection{Synchronisation and measurement description}
This section describes the measurement model of the active and passive sensors and their synchronisation method. It is assumed that the active radar is able to transmit and receive its own waveform in a classical manner, but it emits a special waveform containing the time stamp of the transmission. The passive sensor is also able to receive this waveform and decode it to be able to reconstruct the time stamp.

Both sensors possess their own clocks that can deviate but for simplicity we define the transmitter's clock to be perfect and the error to only be on the receiver's side (master-slave system). The clock drift after $N$ seconds is modelled as a random walk:
\begin{align}
	\varDelta T = \sum_{i=1}^N Z_i,\qquad Z_i \sim \mathcal{U}(-d,d),\quad d > 0.
\end{align}
This defines a discrete version of the drift which is interpolated linearly to model the corresponding continuous extension.
Notice that the error $\varDelta T$ is proportional to the error of the distance measurement from transmitter to target and back to the receiver $\varDelta D$ and thus has a direct impact on the tracking error.
As common trackers expect the measurement to be in spherical coordinates, we transform the original bistatic measurement to a monostatic measurement. Therefore the distance from the transmitter to the target and back to the receiver $D(p)$ is transformed to the range from the target to the receiver $R(p)$ using trigonometric relations.
The corresponding covariance transformation is calculated using the Monte Carlo integration method since it was found out that the linear first order approximation by the Jacobian Matrix leads to instabilities in the Kalman-Filter.
The angular measurement standard deviation is modelled as
\begin{align}
	\sigma_x = 0.628 \, \theta_{x}/(2 \sqrt{\text{SNR}}), \quad x \in \{a,e\},
\end{align}
where $\theta$ gives the \SI{3}{\dB} beamwidth in azimuth and elevation, respectively. The range error standard deviation is assumed to be known and fixed.

\subsection{FHR Cognitive Radar Simulator CoRaSi}

Fraunhofer FHR's Cognitive Radar Simulator (CoRaSi) is a multi-purpose software simulator for phased array radars with electronic beam steering written in Java, which was developed as part of the basic funding by the German Ministry of Defence.
It enables real-time analyses of radar systems in different frequency ranges, rotating or static, with arbitrary antenna patterns and search strategies. The simulation includes the most important functions such as search, tracking, target classification, resource management and data fusion across multiple sensors.
CoRaSi was originally developed for ground-based multifunction radars equipped with AESA antennas,
but is now also capable of simulating ship and air-based systems.
In case more than one sensor is simulated, different fusion concepts (central, decentral, hybrid) are available to merge information.
Due to its variability CoRaSi is able to analyse many phased array radars based on given parameters.

\begin{figure}[t] 
	\centering
	\includegraphics[width=.39\textwidth]{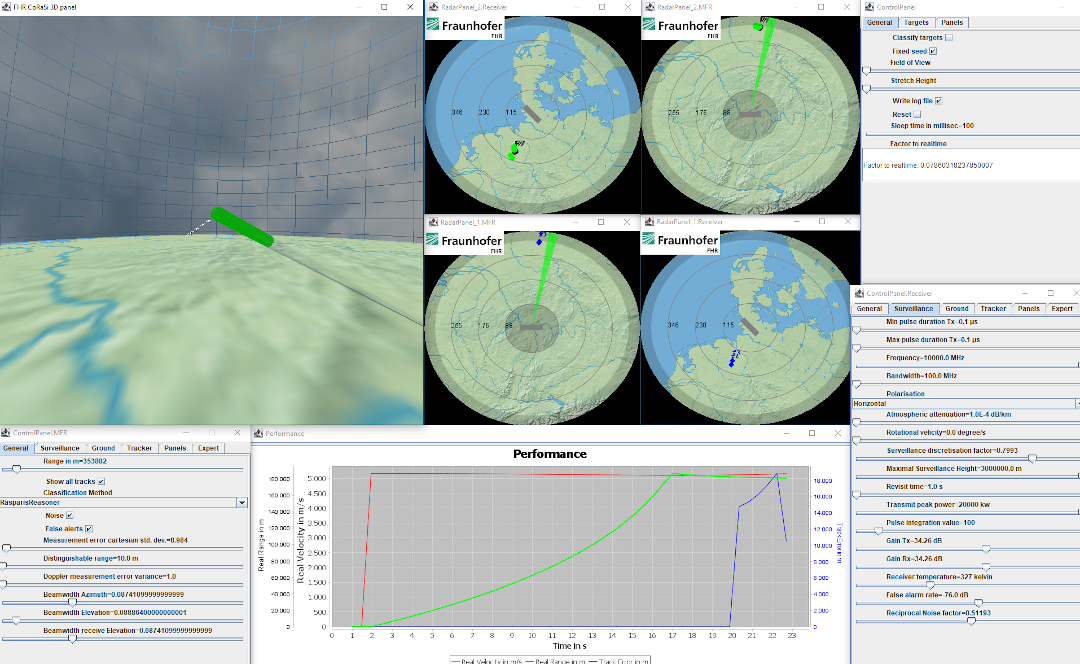}
	\caption{Graphical user interface of CoRaSi.} \label{fig:corasi}
\end{figure}

The simulation also enables the analysis of threat trajectories.
Detections are generated for every target taking into account aspect angle-dependent radar cross-sections.
This includes also the generation of false alarms based on the sensor parameters, which leads to a realistic simulation of target detection.
CoRaSi comes with an optional 3D-GUI allowing for run-time adaptation and evaluation of key parameters as well as an extensive logging scheme for further post-simulation analysis (cf.\ Figure~\ref{fig:corasi}).

\begin{figure}[t] 
	\centering
	\includegraphics[width=.39\textwidth]{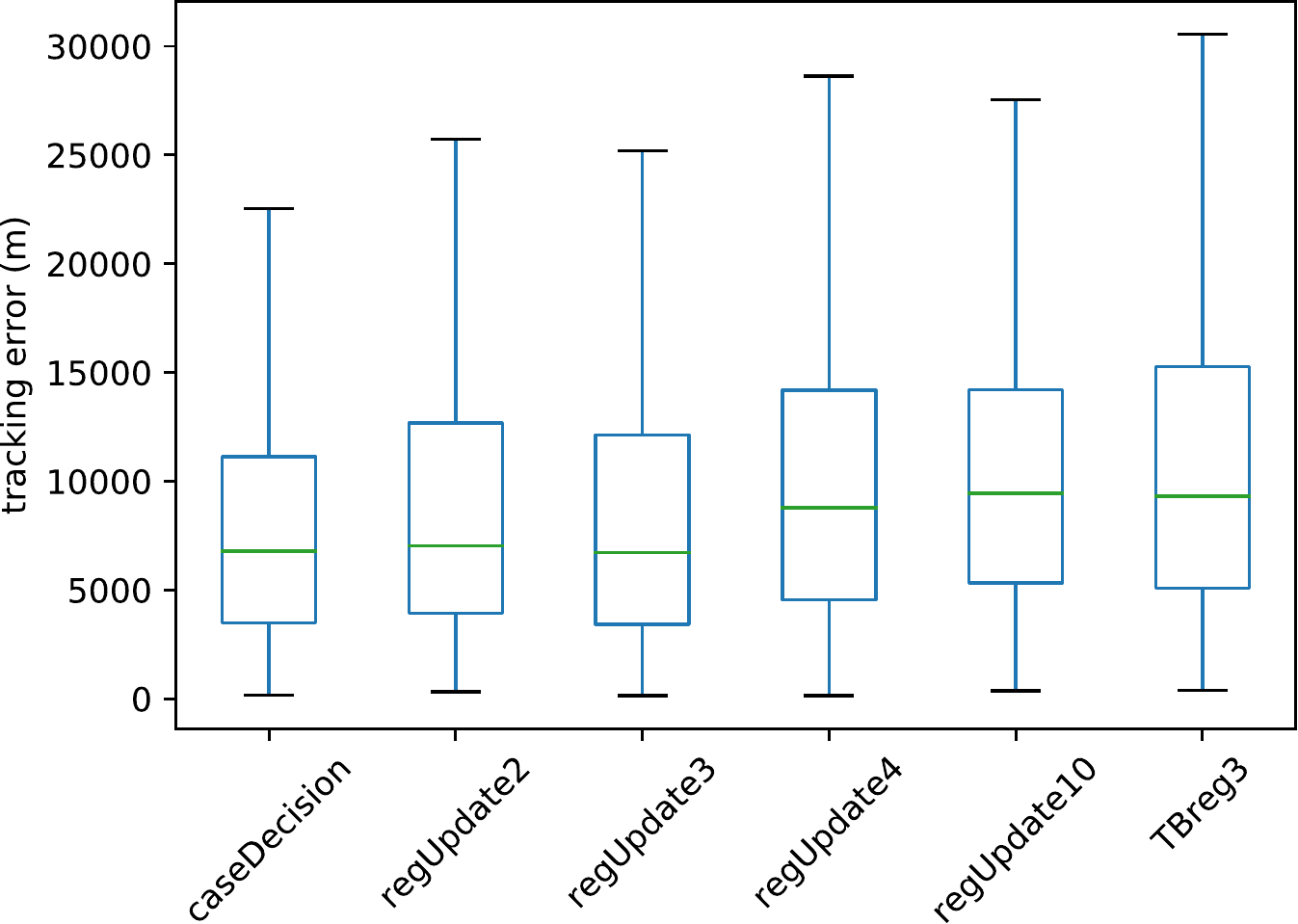}
	\caption{Comparison of tracking performance for various synchronisation schemes. The box plot shows the median (green).
	}
	\label{fig:box}
\end{figure}

\subsection{Experimental results}

We compared the proposed algorithm (called \emph{caseDecision} in tables and figures)  with regular Q-RAM for fixed synchronisation patterns of 1 through 4 and \SI{10}{\s} (called \emph{regUpdate1} etc.) and a rule-based approach with time balanced scheduling (cf.~\cite{butler_tracking_1998}) with a desired synchronisation period of \SI{3}{\s} (called \emph{TBreg3}) in the scenario described above. All Q-RAM based methods make use of a planning interval of length \SI{1}{\s}. For caseDecision, the synchronisation decision is binary. The algorithm can decide whether to perform a synchronisation task with a duration of \SI{231}{\ms} in any given planning interval.
Note that as the system was unable to acquire and track all targets using a fixed synchronisation interval of \SI{1}{\s}, regUpdate1 is not further considered in the following.

Our algorithm performed best with a median track error of \SI{6726.4}{\m} versus \SI{6954.9}{\m} for the next best solution. The mean error on the other hand is improved by \SI{15}{\%} (\SI{8121.5}{\m} vs.\ \SI{9520.8}{\m}), which is a significant improvement.
Furthermore, our algorithm showed the most consistent tracking performance with the lowest standard deviation of the tracking error by far.
Also, the maximum track error was improved by \SI{22}{\%} on average as compared with the next best performance.
Not unexpectedly, the remaining Q-RAM based methods outperformed the rule-based approach TBreg3.
The detailed results are given in Table~\ref{tab:results} and Figure~\ref{fig:box}.

\begin{table}
	\centering
	\caption{Key data of tracking errors per allocation strategy. Results are averaged over 5 simulation runs.}\label{tab:results}
	\begin{tabular}{lrrrrr}
		\toprule
		{} &  median &   min &     max &    mean &  std. dev. \\
		\midrule
		caseDecision &  6726.4 & 410.3 & 29711.9 &  8121.5 &     5758.2 \\
		regUpdate2   &  7526.0 & 440.6 & 43131.8 & 10185.8 &     8790.4 \\
		regUpdate3   &  6954.9 & 267.8 & 38133.3 &  9520.8 &     8089.9 \\
		regUpdate4   &  8452.8 & 367.7 & 44974.2 & 11362.2 &     9382.6 \\
		regUpdate10  &  9299.9 & 671.0 & 42533.1 & 11273.5 &     8036.7 \\
		TBreg3       &  9942.7 & 636.5 & 44276.6 & 12151.2 &     9149.9 \\
		\bottomrule
	\end{tabular}
\end{table}

\begin{figure}[t] 
	\centering
	\includegraphics[width=.45\textwidth]{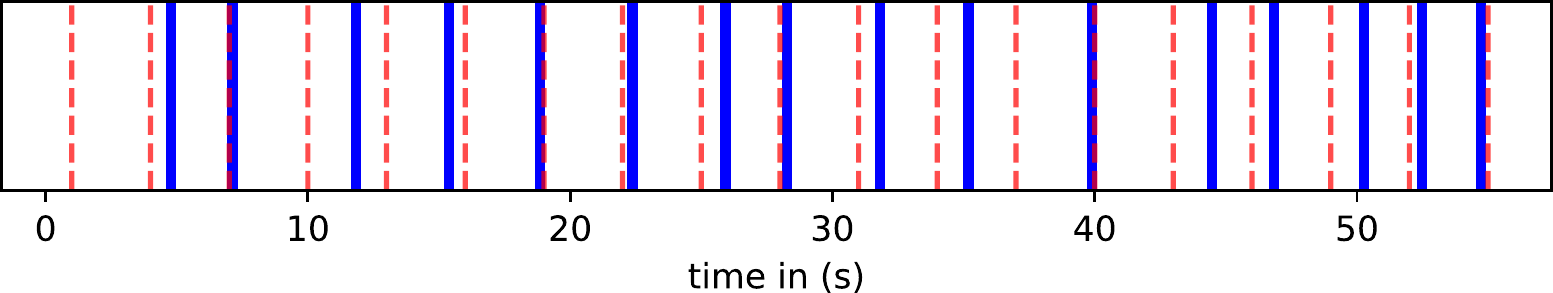}
	\caption{Synchronisation schedules in a single run of the simulation: regular scheme (red) and adaptive scheme (blue).}
	 \label{fig:sync_schedule}
\end{figure}

The adaptive algorithm leads to flexible synchronisation schedules with an improved load balancing and reduced tracking errors. One such resulting schedule is shown in Figure~\ref{fig:sync_schedule}.

\section{Conclusion}
\label{sec:conclusion}

An extension of a Q-RAM based radar resource management architecture has been presented enabling radar systems to cleverly schedule synchronisation tasks to ensure an optimal overall system performance.
Moreover, it has been indicated that the proposed algorithm is applicable in settings other than radar synchronisation.
The provided experimental results demonstrate the effectiveness of the adaptive synchronisation scheme, which significantly outperforms more traditional synchronisation concepts (with regular update rates) in simulations.
Future investigations concern the implementation for
sub nanosecond phase synchronisation
in coherent distributed MIMO networks.
Overall, the presented approach is a valuable step on the way to a truly cognitive radar system.


\bibliographystyle{IEEEtran}
\bibliography{IEEEabrv,lit}

\end{document}